\documentclass[aps,pra,twocolumn,superscriptaddress,groupedaddress]{revtex4}  
\usepackage{graphicx}  % needed for figures
\usepackage{dcolumn}   % needed for some tables
\usepackage{bm}        % for math
\usepackage{amssymb}   % for math
\usepackage{amsmath}
\usepackage[caption = false]{subfig}
\usepackage[justification = raggedright]{caption}

\begin{document}

\title{Block entanglement and fluctuations in finite size correlated electron systems}
\author{Archak Purkayastha}
\email{archakp@icts.res.in}
\affiliation{International Centre for Theoretical Sciences, Tata Institute of Fundamental Research,  Bangalore-560012, India}
\author{V. Subrahmanyam}
\email{vmani@iitk.ac.in}
\affiliation{Department of Physics, Indian Institute of Technology, Kanpur-208 016, India}      
\date{\today}

\begin{abstract}
The block entanglement entropy and fluctuations are investigated in one dimension in finite size correlated electron systems using the Gutzwiller wave function as a prototype correlated electron state. Entanglement entropy shows logarithmic divergence for all values of the correlation projection parameter $g$, as predicted by conformal field theories for critical systems, but the central charge requires finite size corrections. There is an infinite correlation length corresponding to correlation between same kinds of spins, for all values of $g$. A scaling form for the block entropy, as a function of $g$ and the system size $N$, is proposed which predicts a metal-insulator crossover at $N^{1/3} g\approx 0.24$. Bipartite fluctuations in the number of particles in a block, and the spin fluctuations also obey an approximate scaling. A relation is found between the block entropy and the bipartite spin fluctuations.  Our results show some correspondence with an experiment on Ni nanochains.

\vskip 0.5cm
\hfill {PACS: 03.65.Ud, 71.27.+a, 71.30.+h  }
\end{abstract}

\maketitle

\section{\label{Intro} Introduction}

Quantum entanglement of a system, perceived as   a resource in quantum information and communication protocols, quantifies the correlations between the parts of the system\cite{Nielsen}. The entanglement entropy is a widely-used  entanglement measure, along with many other measures, that
has been used to investigate quantum phase transitions in spin systems\cite{plenio}. In particular, in the
vicinity of critical points, the spin correlation functions exhibit long-range behaviour, which reflects on
the behaviour of the entanglement. Thus, it is interesting to study the well-known models of interacting electron/spin systems from the entanglement perspective\cite{sahoo}. In this article, we study the block entanglement for correlated electron states at half filling and for other filling factors.  

Interacting electron systems have more structure than interacting spin qubit systems in the way of more
states per site. We can take them to have two qubits per site, a qubit each for the charge and the spin
degrees of freedom. This severely restricts numerical investigation to studying a finite system with a substantially-fewer number of sites than that of  spin systems. For a system of $N$ sites, the Hilbert space dimension varies from  $4^N$ for uncorrelated electrons to $2^N$ for strongly-correlated electrons (corresponding to infinite repulsion between electrons). In this article, we study the strong correlation effect on the entanglement entropy for the one-dimensional Gutzwiller state\cite{Gutzwiller}. In this state,  the strong on-site correlation effect of the Hubbard model ground state, 
is implemented  through the action of a  projection operator on the non-interacting metallic state.
The metallic state,
vis. the Fermi ground state which is constructed from occupying lowest-lying one-electron states for
both up and down spin electrons, has no correlation between up and down spin electrons. Viewed from
occupation of site-basis states, it has states with doubly-occupied sites with largest probability. The projection operator gives a weighted amplitude to  a basis state
with doubly-occupied sites,  so that the doubly-occupied sites occur with a probability weight that is determined by the correlation effect.

\section*{Block entanglement}

The entanglement in a bipartite system is 
defined as the von Neumann entropy of bipartition. Consider a composite system comprised of two
independent parts, say $A$ and $B$.  The entanglement between $A$ and $B$ in a pure state of the composite system is given by von Neumann entropy of the part $ A$ (or equivalently of part $B$), given by
\begin{equation}
S = - Tr (\rho_A log \rho_A) \label{E},
\end{equation} 
where $\rho_A$ is the reduced density matrix of $A$ obtained by a partial trace over the Hilbert space
of $B$. Since, the reduced density matrix of the subsystem may not correspond to a pure state, though
the parent state is a pure state, it can give rise to an entropy for the subsystem, thus a possibility of
the quantum entanglement arises. The amount of this entanglement entropy itself is a measure of
the quantum correlations between the two subsystems.

Scaling of entanglement entropy with the size of the subsystem has been an important tool to explore the role of quantum correlations in many-body systems. Unlike, the physical observables such as
the internal energy, the entropy and the magnetization which are extensive quantities that are proportional to the volume of the system, the entanglement entropy, along with many other measures of entanglement,  is not extensive.  In this connection the area law
has been proposed\cite{srednicki, EMB} which says  that the entanglement entropy scales as the area of the surface separating the two subsystems. In one dimension, this means it is a constant, as the surface consists of just one point. The understanding behind the area law is that for systems with short-ranged interactions, the quantum correlations between the two subsystems should occur close to the boundary surface separating the two subsystems. However, area law is modified by
a logrithimic correction for gapless spin systems, which are known to be in a critical regime\cite{plenio}. This is because in the critical regime, the correlation length diverges, and correlations remain non-zero  even at very large length scales. For one dimension,  an explicit result has been obtained for gapless spin systems using conformal field theory (CFT)  which describes the continuum limit for such systems \cite{CFT2}. The result relates the coefficient of logarithmic correction to the central charge of the theory.
For a system of $N$ sites the entanglement entropy, between part $A$ consisting of a block of $L$ contiguous sites and the part $B$ containing the rest of the sites, shows a universal behavior, given by
\begin{equation}
S= \frac{c}{3}\log_2{L} + c_1 \label{EEspin}
\end{equation}
where $c$ is the central charge of the underlying CFT, and $c_1$ is a non-universal constant of  $O(1)$.

The success of entanglement entropy scaling has lead to various proposals for ways to measure it experimentally \cite{measure1,measure2,measure3}. With these, entanglement entropy has ceased to be a merely theoretical tool. We will see later how the entanglement scaling can be used to infer a metal-insulator transition in finite-sized chains of correlated electrons.

The standard technique for calculating entanglement entropy for a pure state is through the Schmidt decomposition\cite{Nielsen} of a bipartite state.  An arbitrary pure bipartite state can be written, using
direct product basis states of $A$ and $B$, as
\begin{equation}
| \psi \rangle = \sum_{i,j} ~\gamma_{i,j} ~|i\rangle_A ~| j \rangle_B
\end{equation}
where $i (j)$ labels an orthonormal basis of the part $A(B)$. Let the dimension of the Hilbert space
of $A(B)$ be $D_A (D_B)$, and let $D_A<D_B$. We can view the  amplitude $\gamma_{i,j}$ as an element of a matrix $\hat \gamma$ of dimension $D_A$ by $D_B$. Though there are $D_A D_B$ number of wave function amplitudes in the above state,  the above state can be brought to the Schmidt
form  containing  atmost $D_A$ terms. On first performing the independent sum over $j$ in the above for a fixed state $i$ of the part $A$, we get $D_A$ number of states for $B$, labelled by $i$, i.e. $|\tilde i\rangle_B \equiv \sum_j \gamma_{i,j} |j\rangle_B$. These superposition states of $B$ may not form orthonormal basis. After the process
of orthogonalization of these $D_A$ states we get an effective sum, with a square matrix $\hat \gamma_{eff}$ of dimension $D_A$. The Schmidt numbers and  the Schmidt basis states are
obtained from diagonalising the effective matrix. Thus, the above state can be written in a Schmidt-decomposed form, given by

\begin{equation}
|\psi_ \rangle = \sum_{n=1}^{D_A} \sqrt{\lambda_{n}} | n \rangle_A | n \rangle_B,
\end{equation}
where the Schmidt numbers $\lambda_n$ are given by the eigenvalues of $\hat \gamma \hat \gamma^T$, and the
basis states for $A$ and $B$ are linear combinations of the original bases. The maximum number of
non-zero Schmidt numbers is given by $D_A$, and hence the von Neumann entropy of the reduced density matrix for both $A$ and $B$ are equal. It is straightforward to
get the entanglement entropy (as $\lambda_n$ are the eigenvalues of the reduced density matrix 
$\rho_A$),

\begin{equation}
S=  - \sum_n \lambda_n  \log_2 \lambda_n  \label{Eformula}.
\end{equation}
The dependence of the entanglement on the correlation effect is carried by the Schmidt numbers.

Even though the case for spin systems in one dimension has been thoroughly explored via entanglement entropy as well as the entanglement spectrum, the same is not true for correlated electron systems. The goal of the present work is to investigate these measures using the Gutzwiller state as a prototype state of correlated electron systems. We will see later that the entanglement entropy, for
equal bipartition in the Gutzwiller state follows the
logarithmic relation given in Eq.2, for all parameter regime. However at half filling, a size-independent central
charge can be assigned only at the two limits of uncorrelated metallic state at one end (with $c=2$) and
a strongly-correlated insulating state at the other end ($c=1$). In the metal-insulator crossover regime, the coefficient shows a system-size dependence, thus not conforming to a CFT prediction.

\section*{The Gutzwiller state}
The simplest model that has the essential ingredients of  de-localized lattice electrons and a screened Coulomb interaction is the Hubbard model\cite{H}, given as

\begin{equation}
H= t\sum_{\langle ij \rangle \sigma} c_{i\sigma}^\dag c_{j\sigma}  + U\sum_i \hat n_{i\uparrow}\hat n_{i\downarrow}.
\end{equation}

Here,  $c_{i\sigma}^\dag$ creates an electron with the given spin projection at site $i$, and $\hat n_{i\sigma}$ is the corresponding number operator. The first represents the hopping of electrons on nearest-neighbour sites, with an amplitude $t$. The strong correlation (which is just on-site density-density
interaction) arises from the second term, where $U$ is the interaction strength.
Even though analytical solution exists in one dimension, a theoretical treatment is made very tedious by the fact that the kinetic term is diagonal in momentum basis whereas the interaction term is diagonal in site basis. A variational approach to the problem was proposed by Gutzwiller \cite{Gutzwiller}, suggesting a trial ground state where correlations were introduced into the non-interacting metallic state by a local correlation factor in site basis\cite{metzner}. The Gutzwiller state is written, for a lattice of $N$ sites, as
\begin{equation}
| g\rangle = \prod_{i=1}^N\{1-(1-g)\hat{n}_{i\uparrow}\hat{n}_{i\downarrow}\}|F\rangle,
\end{equation}

where the  metallic Fermi state $|F\rangle =\prod_{k=0}^{k_F} \hat{c}_{k\uparrow}^\dagger \hat{c}_{k\downarrow}^\dagger |0\rangle$ is constructed from the vacuum state by using products of momentum space operators  $\hat{c}_{k\sigma}^\dagger$. Here, 
$k_F$ is the Fermi momentum which is controlled by the filling factor. 

 The correlation projection parameter $ g$ varies from 0 to 1,  incorporating the correlation effect of the Hubbard-$U$.  The insulating limit corresponds to $g\rightarrow 0$ or  $U\rightarrow \infty $ and the
 metallic limit is $g\rightarrow 1$ or $U\rightarrow 0$. At half-filling, the two limits describe a change from completely de-localised state ($g=1$) to completely localised state ($g=0$), giving rise to a metal to insulator transition. 

An exact calculation by Metzner and Vollhardt \cite{metzner} gives a relation between $g$ and $t/U$,  at half-filling in one dimension, we have,
\begin{equation}
g \hspace{2pt} \log{g} = -\frac{4t}{\pi U} \label{gU}
\end{equation}

Gutzwiller state provides a simple example of a Fermi liquid\cite{FL}. Certain properties of Fermi liquid state of transition metals calculated from the Gutzwiller state have been compared with experiments. 
The properties of Ni \hspace{4pt}\cite{Ni} as well as of normal liquid $^3$He \hspace{4pt}\cite{He3} which can be treated as a Fermi liquid, have been successfully described by the Gutzwiller
state. Fermi liquids are characterised by a sharp Fermi surface, i.e. a discontinuity at the Fermi level in the single-particle momentum distribution. However, in one dimension Fermi liquids usually break down into Luttinger liquids which do not have the discontinuity at the Fermi level. 

 Interestingly, this discontinuity and fermi-liquid behavior have been seen to occur in some finite-size systems even in one dimension\cite{1D}. 
 The Gutzwiller state shows the discontinuity at the Fermi level for non-zero $g$, thus differing in this respect from the physics of one-dimensional Hubbard model.  This Fermi surface discontinuity for a one-dimensional system at half-filling  is given by\cite{metzner},
\begin{equation}
Z(g) = \frac{4g}{(1+g)^2} \label{discont}.
\end{equation}
At $g=0$, the discontinuity disappears and the system passes to an insulating state (Brinkman-Rice insulator\cite{BR}). Near the insulating regime, another quantity of interest which goes to zero is the average density of doubly-occupied sites, which is  given in one dimension by\cite{metzner}
\begin{equation}
d(g) = \frac{g^2}{2(1-g^2)^2}[-n(1-g^2)-\log \{1-n(1-g^2)\}]\label{double}
\end{equation}
where $n$ is the number density of electrons.

Though the Gutzwiller state is an well-explored state and many exact results are known, entanglement properties of the state remain fairly unexplored. In a previous work,  the global entanglement of Gutzwiller state, which measures the average site purity in the state, was investigated\cite{A}. It was seen to be directly related to the  average double occupancy of Gutzwiller state. The entanglement for the strongly-correlated case was seen to be smaller than that of the uncorrelated metallic state for most filling factors. This was mainly due
to the fact that a strong correlation (a large $U$) enhances diagonal spin correlation functions,  whereas usually entanglement is enhanced by off-diagonal correlations and decreased by
diagonal correlations, as seen in Heisenberg 
antiferromagnets\cite{subrah1}.  In this work, we investigate the entanglement entropy and entanglement spectrum of the Gutzwiller state in one dimension.

\begin{figure}[t]

\includegraphics[scale=0.45] {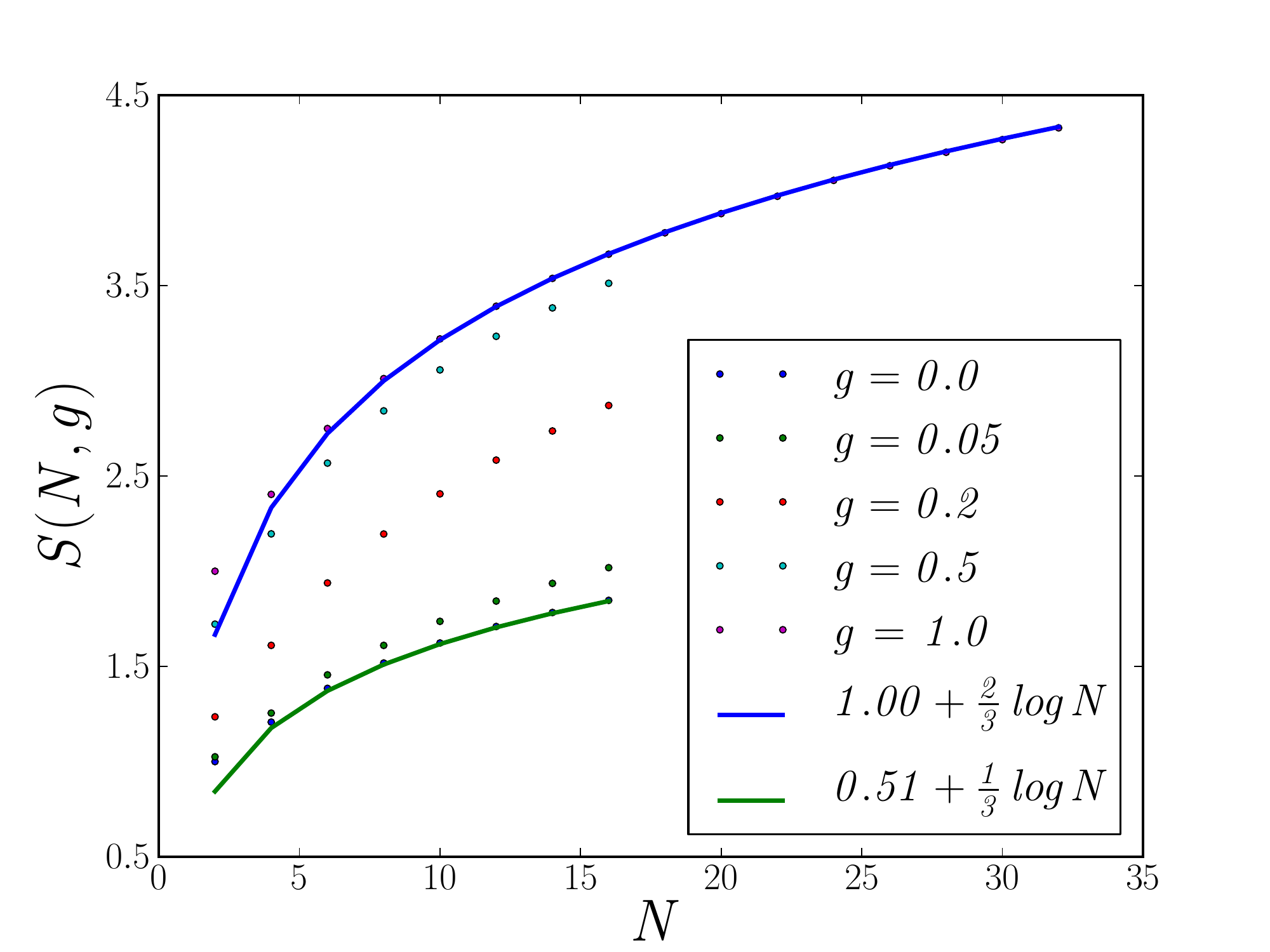}
\caption{\label{fig:EvsN} Entanglement entropy $S(N,g)$ as a function of the number of sites $N$ is shown for various values of on-site correlation factor $g$ at half-filling. Continuous lines show verification of CFT results for the metallic and the insulating limits, given in Eq.\ref{c}.  } 
\end{figure}

\section{\label{EEGS}Entanglement entropy of Gutzwiller state}
\begin{figure*}

\begin{tabular}{cccc} 
\subfloat[\label{fig:evsy}]{\includegraphics[scale = 0.45]{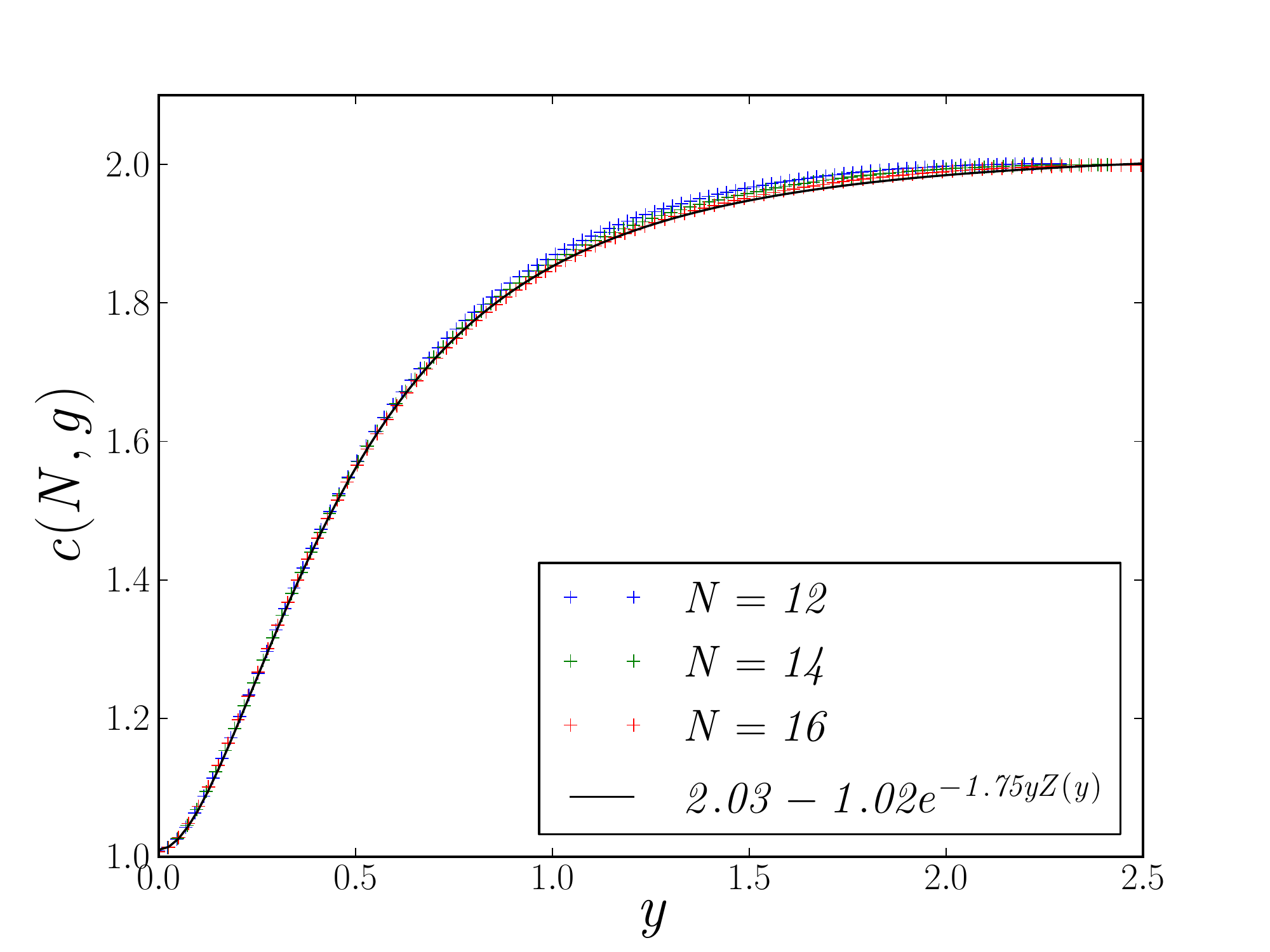}} &
\subfloat[\label{fig:de/dy}]{\includegraphics[scale=0.45]{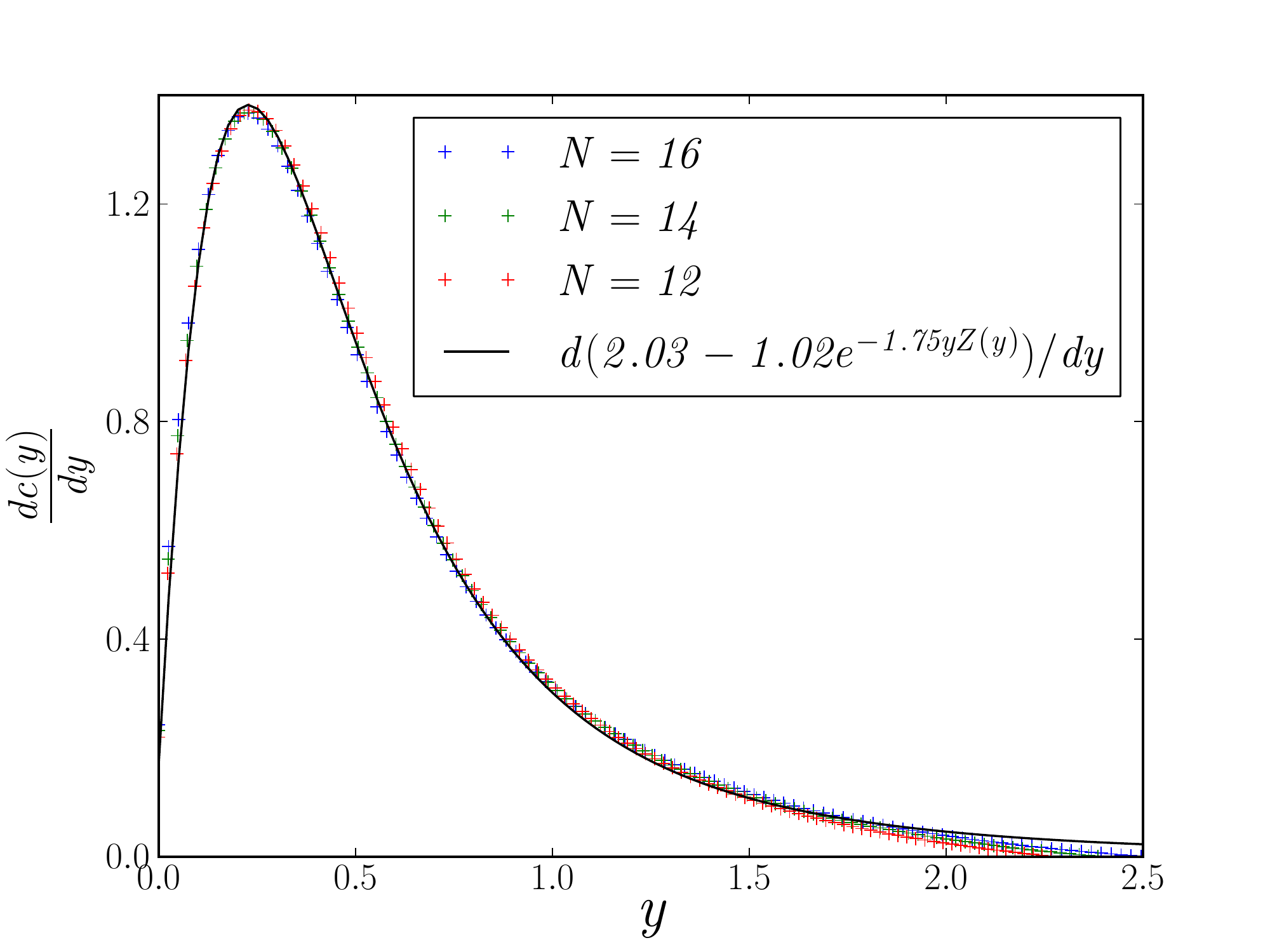}} &

\end{tabular}
\caption{(a) The scaling function $c(N,g) =S(N,g)/S(N,0)$ as a function of the scaling variable $y = N^{\frac{1}{3}} g$ at half-filling, showing a data collapse on to a single curve. 
We also show a fit for $c(y)$, (Eq.\ref{E(y)}) where $Z(y)$ is the function for the  discontinuity in single-particle momentum distribution of electrons at Fermi level defined by Eq.\ref{discont}. (b) The derivative of the scaled entanglement entropy $\frac{dc(y)}{dy}$ is plotted with the scaling variable $y = N^{\frac{1}{3}} g$ at half-filling. The peak at $y=y_0 \simeq 0.24$ signifies the metal-insulator crossover region. The figure also shows the derivative of the fit given in Eq.\ref{E(y)}. The fit is not exact but describes the peak quite well. } 
\end{figure*}

We consider a closed chain of $N$ sites and a paramagnetic case where there is no net magnetization. There are $N_e$ electrons with equal numbers up and down spin polarizations,
$N_{\uparrow}=N_{\downarrow}= \frac{N_e}{2}$.

The bipartition is done in two blocks of equal size, the part $A$ containing the first half of the lattice, and the rest of the lattice forming the part $B$. Therefore, the length of subsystem is 
$L = \frac{N}{2}$. Such bipartition maximizes entanglement entropy. As explained in the previous section, we need to calculate von-Neumann entropy of subsystem, which gives
the entanglement entropy $S(N,g)$, as a function of the system size $N$ and the correlation projection parameter $g$. Numerical calculation of the eigenvalues of the reduced density matrix $\rho_A$
is still quite difficult due to exponentially growing size of the Hilbert space of the block $A$ for a general value of $g$.

Even though is it difficult to calculate entanglement entropy in Gutzwiller state for arbitrary on-site correlation factor $g$, 
the limiting case of the uncorrelated state at $g=1$ is quite easy to manage, as the system reduces to a free fermion state. For this state, the eigenvalues of reduced density matrix can be generated from those of the single particle density matrix of particles with one kind of spin, and the size of such single particle density matrix increases only linearly with system size\cite{Ingo}. 

We have used this technique to verify the scaling relation for the entanglement entropy given in Eq.2 for $g=1$ case. For an arbitrary $g$, entanglement entropy has been calculated numerically using the method of Schmidt decomposition discussed earlier. Within our computational limitations, we could go upto 32 sites for $g=1$ and upto 16 sites for all other values of $g$.

\section*{Half-filled Gutzwiller state, $N_{\uparrow}=N_{\downarrow}= \frac{N}{2}$}
We first consider the half-filled Gutzwiller state ($N_e=N$) with zero magnetization ($N_\uparrow=N_\downarrow$).
  
 At half-filling, in $g=0$ limit, the Gutzwiller state reduces to ground state of a Heisenberg-like spin model, whereas in $g=1$ limit, it becomes ground state of a free fermion model. Continuum limits of both the limiting cases are amenable to the  CFT methods\cite{CFT4}. The $g=0$ case is described by a CFT with central charge $c=1$ (as there are no charge degrees of freedom in this case),  and $g=1$ case by $c=2$ (a
unit of central charge each for the spin and charge degrees of freedom in this limit). The corresponding results for the entanglement entropy conform to Eq.\ref{EEspin}, we have,
 \begin{equation}
 \label{c}
 \begin{split}
 S(N,0) &= c_1 + \frac{1}{3}\log_2 N 
 \\
 S(N,1) &= c_2 + \frac{2}{3}\log_2 N 
 \end{split}   
\end{equation}
In both these limits, the system is gapless, and thus exhibits critical behavior.  

Fig.\ref{fig:EvsN} shows variation of $S(N,g)$ with $N$ for different values of $g$. The plots for $g=1$ and $g=0$ cases show that already at a size of 8 sites, the results match quite well with the the CFT form. This is a clear sign of scale invariance in these two limits. However, we see that entanglement entropy increases with system size for all values of $g$. This means the area law is violated always, indicating a crossover critical behaviour from the insulating Heisenberg spin limit at $g=0$ to the uncorrelated metallic limit at $g=1$.

The crossover phenomena are best understood in terms of a scaling form. Using the fact  that $S(N,1)$ is nearly double of $S(N,0)$ and that $S(N,g)$ increases with $N$ for all values of $g$, we propose that  a scaling form should emerge for the scaled entanglement $c(N,g)$, given by
\begin{equation}
c(N,g)= \frac{S(N,g)}{S(N,0)}=\frac{S(N,g)}{\frac{1}{2}+\frac{1}{3}log_2 N} = c(y) \label{e(y)1}
\end{equation} 
where $y=y(N,g)$ is the scaling variable. $c(N,g)$ is nothing but the central charge at the corresponding $g$ with corrections due to finite size $N$. For a scaling form to emerge, $c(y)$ must satisfy the following conditions :
\begin{align}
c(y(N,g=0)) \simeq 1 & \hspace{8pt}  \label{condn1} \\
c(y,(N,g=1)) \simeq 2 & \hspace{8pt}  \label{condn2} \\
c(y(N\rightarrow \infty, g\neq0)) \simeq 2 \label{condn3}
\end{align} 

Eq \ref{condn1} and Eq \ref{condn2} are just restatement of the CFT results in the insulating and metallic limits respectively. The justification of Eq. \ref{condn3} is that, it is known from Metzner-Vollhardt's paper \cite{metzner}, in the thermodynamic limit, half-filled Gutzwiller state is insulating only exactly at $g=0$, and metallic otherwise.  So, for a scaling form to emerge, scaling variable $y(N,g)$ must be such that, given Eq \ref{condn1} and Eq \ref{condn2}, Eq \ref{condn3} is automatically satisfied without imposing any new condition on $c(y)$. Also, we expect $y(N,g)$ to be a smooth function. With a little thought, it can be seen that to satisfy these conditions, a general form of $y(N,g)$ should be a sum of products of increasing functions of $N$ with increasing functions of $g$ provided the functions of $g$ are zero at $g$ equal to zero. To check this, we see that for such definition of $y(N,g)$, as $N\rightarrow\infty$, $y\rightarrow\infty$ for all non-zero $g$. This means all non-zero values of $g$ give same value of $c(y)$ which, therefore, has to be the value at $g=1$. At $g=0$, $y=0$ for all $N$, thereby giving the correct value of $c(y)$ even as $N\rightarrow\infty$.  Assuming each of these functions can be expanded in a power series, to the leading order it must be linear in $g$, multiplied by some increasing function of $N$. Assuming a simple form $N^p$ for the function of $N$ gives :
\begin{equation}
y \simeq N^p g,  \hspace*{10pt} p>0. \label{y}
\end{equation}

\begin{figure*}

\begin{tabular}{cccc}
\subfloat{\includegraphics[scale = 0.45]{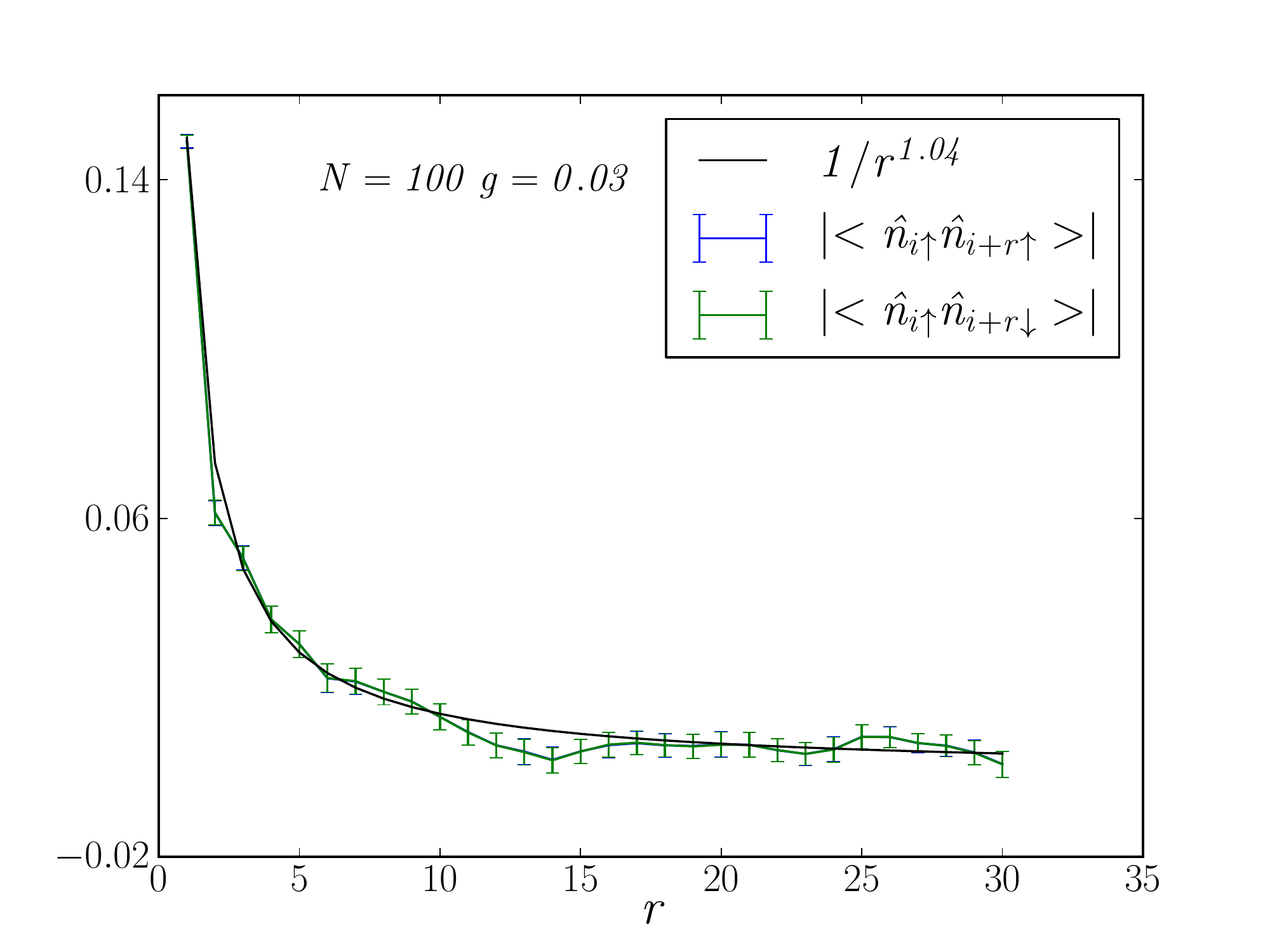}} & 
\subfloat{\includegraphics[scale = 0.45]{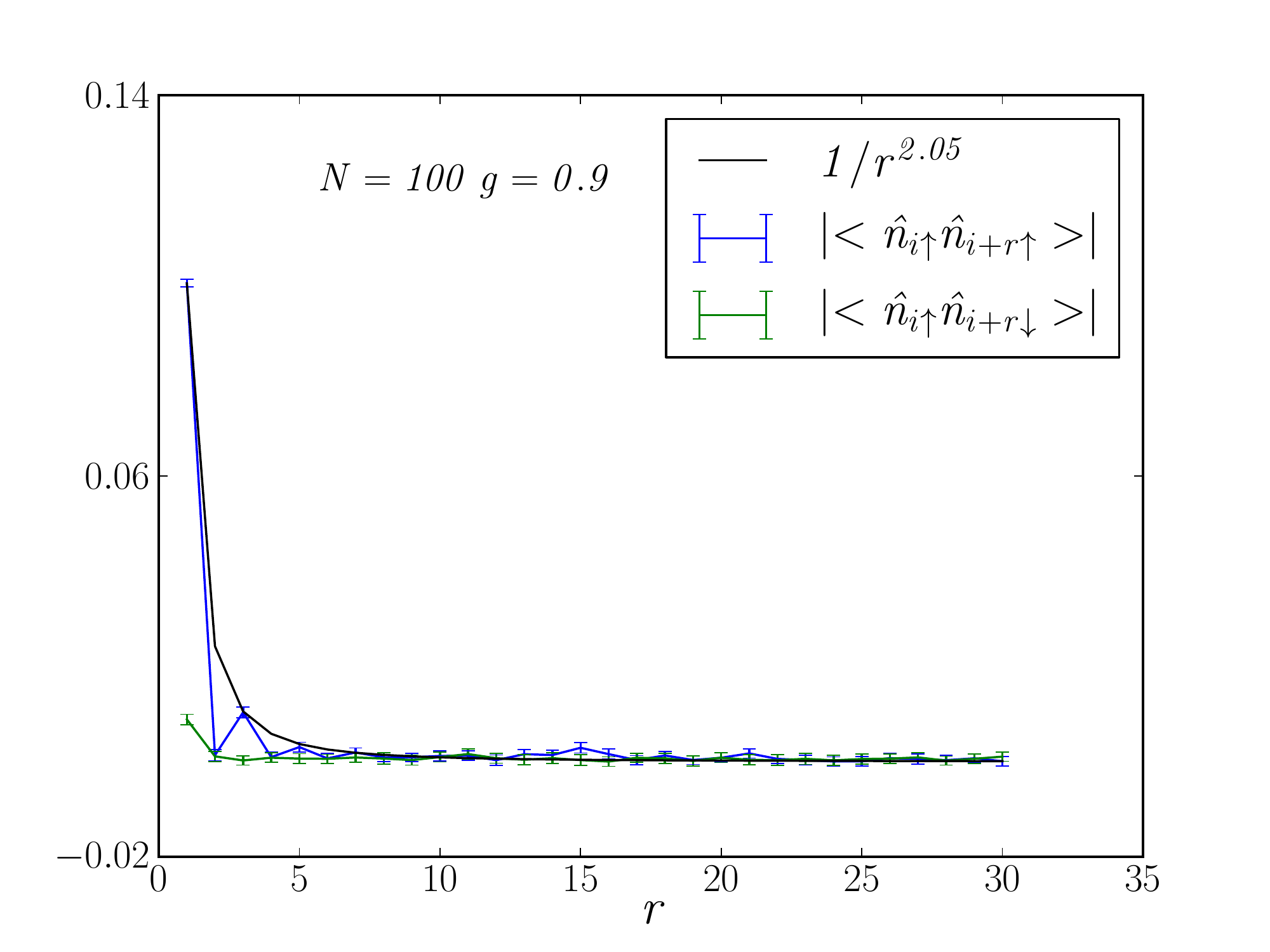}} &

\end{tabular}
\caption{\label{fig:corrvsr} Correlation functions $\mid<\hat{n}_{i\uparrow} \hat{n}_{i+r\uparrow}>\mid$ and $\mid<\hat{n}_{i\uparrow} \hat{n}_{i+r\downarrow}>\mid$ are plotted with $r$, and the corresponding fits of $\mid<\hat{n}_{i\uparrow} \hat{n}_{i+r\uparrow}>\mid$ to a power law are shown for $g=0.03$ and $g=0.9$. The power law suggests an infinite correlation length for $\mid<\hat{n}_{i\uparrow} \hat{n}_{i+r\uparrow}>\mid$ over the entire range of $g$. $\mid<\hat{n}_{i\uparrow} \hat{n}_{i+r\downarrow}>\mid$ is seen to overlap with $\mid<\hat{n}_{i\uparrow} \hat{n}_{i+r\uparrow}>\mid$ for $g=0.03$, thereby showing a power law decay, but is seen to decay exponentially for $g=0.9$. This gives direct verification of our qualitative understanding from entanglement calculations. } 
\end{figure*}

\begin{figure}[t]
\includegraphics[scale=0.45]{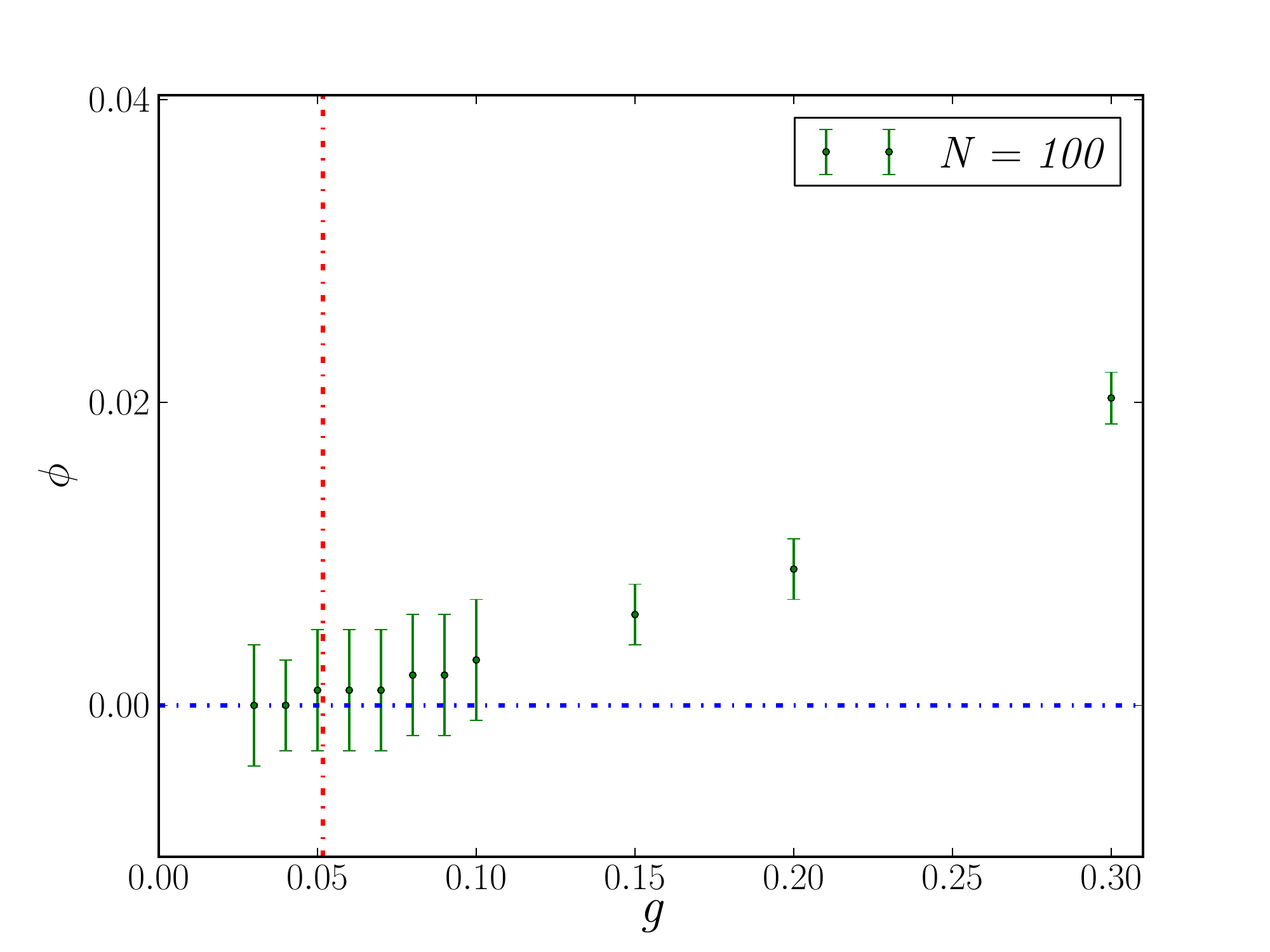}
\caption{\label{fig:mvsg} $\phi = \mid<\hat{n}_{i\uparrow} \hat{n}_{i+1\uparrow}>\mid - \mid<\hat{n}_{i\uparrow} \hat{n}_{i+1\downarrow}>\mid$ is plotted as a function of $g$ for $N=100$. The vertical line shows the position of  $g_c \simeq \frac{0.24}{N^{1/3}}$. The plot shows that $\phi$ starts taking non-zero values as $g\rightarrow g_c$, and beyond that, $\phi$ keeps on increasing with $g$. These observations validate our understanding from entanglement calculations.    }
\end{figure} 

If our claim is true, plots of $c(y)$ vs $y$ must show a data collapse. It is seen that a data collapse (see Fig.\ref{fig:evsy}) is obtained for $p=\frac {1}{3}$ which is quite good upto a certain value of $y$.  There is a small mismatch beyond two significant figures for intermediate values of $y$. This mismatch decreases with $N$. This mismatch is clearly a finite size effect, and stems from our choice of $y$. Since exact value of entanglement has no physical significance, and only the variation of entanglement matters, we will neglect this small mismatch for the time being. Therefore there is an approximate scaling in terms of $y$ which is especially valid if $y$ is not too big.

A more interesting thing to look at is the derivative of $c(y)$ with respect to $y$ (Fig \ref{fig:de/dy}). This is also quite good data collapsed upto a certain value of $y$ and shows dispersion for higher values. Most interestingly, there occurs a peak in the data collapsed region at $y=y_0\simeq 0.24$. Below $y_0$, $c(y)$ increases at an increasing rate with $y$. But beyond $y_0$, it increases at a decreasing rate. Hence, saturation to metallic limit begins at $y_0$. Thus, $y_0$ is the insulator to metal crossover point. Since close to $g=0$ and to $g=1$, the central charge should be nearly independent of $N$, the derivative tends to zero in these two limits. In between, the larger value of derivative is due to finite size effects. Therefore, $y_0$ marks the point where finite size effect is maximum.   

Next, we endeavour to find an analytic form for the data collapsed curve. Motivated by the nature of the numerical curve, we try an exponential function of the form: $ \alpha - \beta e^{-\kappa f(y)}$, where $f(y)$ is an increasing function of $y$, and $\alpha$,$\beta$,$\kappa$ are constants to be determined numerically. From the work of Ding, Seidel, and Yang \cite{FL2}, it is known that entanglement entropy of Fermi liquids depend on the gap at the Fermi level. Since in the Gutzwiller state, the Fermi liquid like property is retained even in one dimension, we expect $f(y)$ to depend in some way on the gap at the Fermi level. By trial, we find that $f(y) = y Z(y)$, where $Z(y)$ is the expression for gap at the Fermi level given by Eq.\ref{discont}, gives a good fit to the computational result. So we have,

\begin{equation}
c(y)\simeq \alpha - \beta e^{-\kappa yZ(y)} \label{E(y)}
\end{equation}
where the numerical coefficients are chosen to be,

\begin{equation}
\alpha \simeq 2.03, \hspace{8pt},
\beta \simeq 1.02, \hspace{8pt}, 
\kappa \simeq 1.75,
\end{equation}
 
The above analytic form is clearly not exact. But from Fig \ref{fig:evsy}, is it seen to be almost exact for the range of $y$ where the data collapse is also very good. Especially, from plot of the derivative in Fig \ref{fig:de/dy}, the above function is seen to describe the peak at $y_0$ quite well.

The fact that area law for entanglement entropy is violated for all values of $g$, but still there is a finite-size effect implies there is an interplay of two different length scales in the problem. The two length scales are the two correlation lengths associated with the correlation between similar kind of spins, and correlations between different kinds of spins. The observed results can be explained in terms of these correlation lengths as follows. When CFT results are valid, they imply conformal invariance, and thereby infinite correlation lengths. In the present case, the CFT result for entanglement entropy is valid at $g=0$ and $g=1$. In between, entanglement entropy is described by a similar function but only with finite size corrections to the central charge. Since area law of entanglement entropy is violated for all values of $g$, there is always a correlation length which is infinite. This correlation length is that associated with correlation between spins of same kind. This arises from Pauli's exclusion principle and is always present for any value of $g$. The correlation between spins of different kinds is, however, controlled by the correlation factor $g$. It is the correlation length $\xi_c$ associated with this correlation that changes with $g$ and is responsible for the finite size corrections to the central charge. For a system of size $L$, ( for half-filling, $L=N$), the physics is governed by the smaller between $\xi_c$ and $L$, since the other correlation length is always infinite. So, we can claim that the physics is governed by the ratio $\frac{\xi_c}{L}$. For infinitely correlated case at $g=0$, $\frac{\xi_c}{L} \rightarrow \infty$, for the uncorrelated case at $g=1$, $\frac{\xi_c}{L}=0$. These are the values of $\frac{\xi_c}{L}$ when the system is insulating and metallic respectively. Hence there is a crossover from insulating to metallic when $\frac{\xi_c}{L} \sim 1$. This is also the case when finite size corrections to central charge will be maximum. Therefore, this corresponds to $y_0$. As $L \rightarrow \infty$, for any finite value of $\xi_c$, $\frac{\xi_c}{L}=0$. Thus for any $g\neq 0$, the system is metallic, which is the correct description in the thermodynamic limit \cite{metzner}. The qualitative picture that emerges out of this is that for finite size $L$, there is a non-zero value of $g$ where $\frac{\xi_c}{L} \sim 1$ and a metal-insulator crossover occurs, and this value of $g$ decreases with increase in system size until it becomes zero in the thermodynamic limit.

It is important to note that this qualitative understanding does not depend on our choice of $y$, nor on the ansatz Eq \ref{E(y)}. The signatures of the above argument are only the violation of area law and the occurrence of a peak when entanglement entropy is differentiated with respect to $g$, and that the peak occurs at lower values of $g$ for higher $N$. Our choice of $y$ tells us, at least within the range of system sizes we have studied, the value of $g$ at which the peak occurs corresponds to that at $y_0$, and the derivative of ansatz Eq \ref{E(y)} gives a good description of the peak.

\section*{Variational Monte Carlo for N=100}

We can check our understanding from entanglement calculations by direct calculation of connected correlation functions for half-filled Gutzwiller state  by variational Monte Carlo method using Metropolis algorithm. To this end we calculate the connected density correlation functions $<\hat{n}_{i\uparrow} \hat{n}_{i+r\uparrow}>$ and $<\hat{n}_{i\uparrow} \hat{n}_{i+r\downarrow}>$ for $N=100$. The plots of $\mid<\hat{n}_{i\uparrow} \hat{n}_{i+r\uparrow}>\mid$ and $\mid<\hat{n}_{i\uparrow} \hat{n}_{i+r\downarrow}>\mid$ are shown in Fig. \ref{fig:corrvsr} for two typical values of $g$ in strongly correlated and weakly correlated regions. We find that $\mid<\hat{n}_{i\uparrow} \hat{n}_{i+r\uparrow}>\mid$ shows a power law behaviour for all values of $g$, thereby indicating an infinite correlation length for all $g$. On the other hand $\mid<\hat{n}_{i\uparrow} \hat{n}_{i+r\downarrow}>\mid$ exactly overlaps with $\mid<\hat{n}_{i\uparrow} \hat{n}_{i+r\uparrow}>\mid$ in the strongly correlated limit, thereby showing a power law behaviour, but shows exponential decay in the weakly correlated limit. This is exactly as expected from our qualitative understanding.

These correlation functions can be used to track to insulator to metal crossover as $g$ increases. For this, we look at 
\begin{equation}
\label{m}
\phi = \mid<\hat{n}_{i\uparrow} \hat{n}_{i+1\uparrow}>\mid - \mid<\hat{n}_{i\uparrow} \hat{n}_{i+1\downarrow}>\mid
\end{equation}  

In the strongly correlated or `insulating' region, $\phi=0$. In the weakly correlated, or `metallic' region $\phi$ has a finite value. In between, there is the crossover value of $g=g_c$ near which $\phi$ begins to take on finite values. Assuming Eq. \ref{E(y)} is valid, we expect
\begin{equation}
\label{g_c}
g_c \simeq \frac{0.24}{N^{1/3}}
\end{equation} 
 
Plot of $\phi$ vs $g$ is shown in Fig \ref{fig:mvsg} for $N=100$. The plot shows that $\phi=0$ for small values of $g$, $\phi$ starts taking non-zero values as $g\rightarrow g_c$, and beyond that, $\phi$ keeps on increasing as the system `crosses over' to the `metallic' state. These observations validate our understanding for $N=100$.

 \begin{figure}[t]
 \includegraphics[scale=0.45]{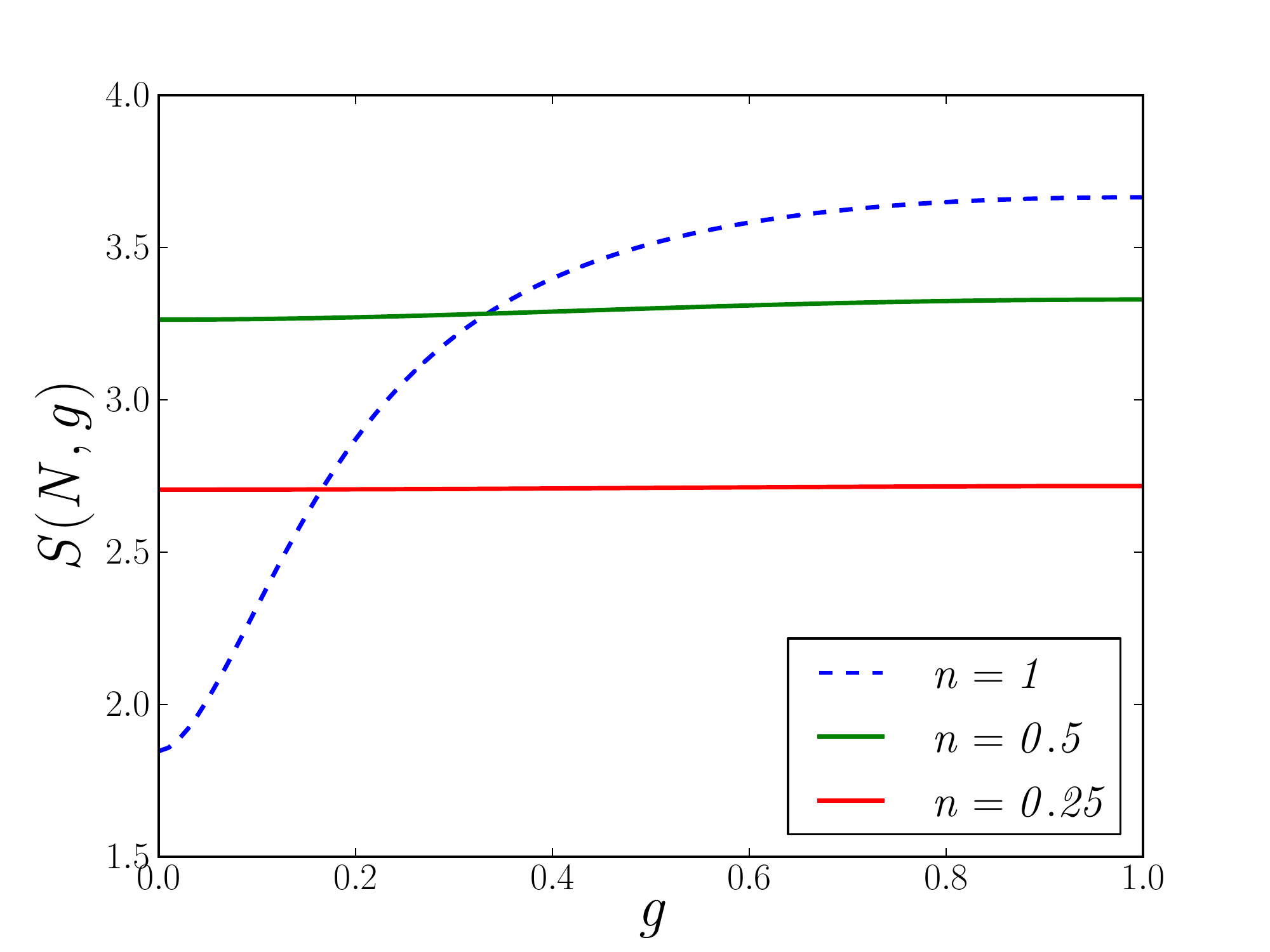}\llap{\makebox[7cm][t]{\raisebox{.7cm}{\includegraphics[width=3.5cm,height=2.3cm]{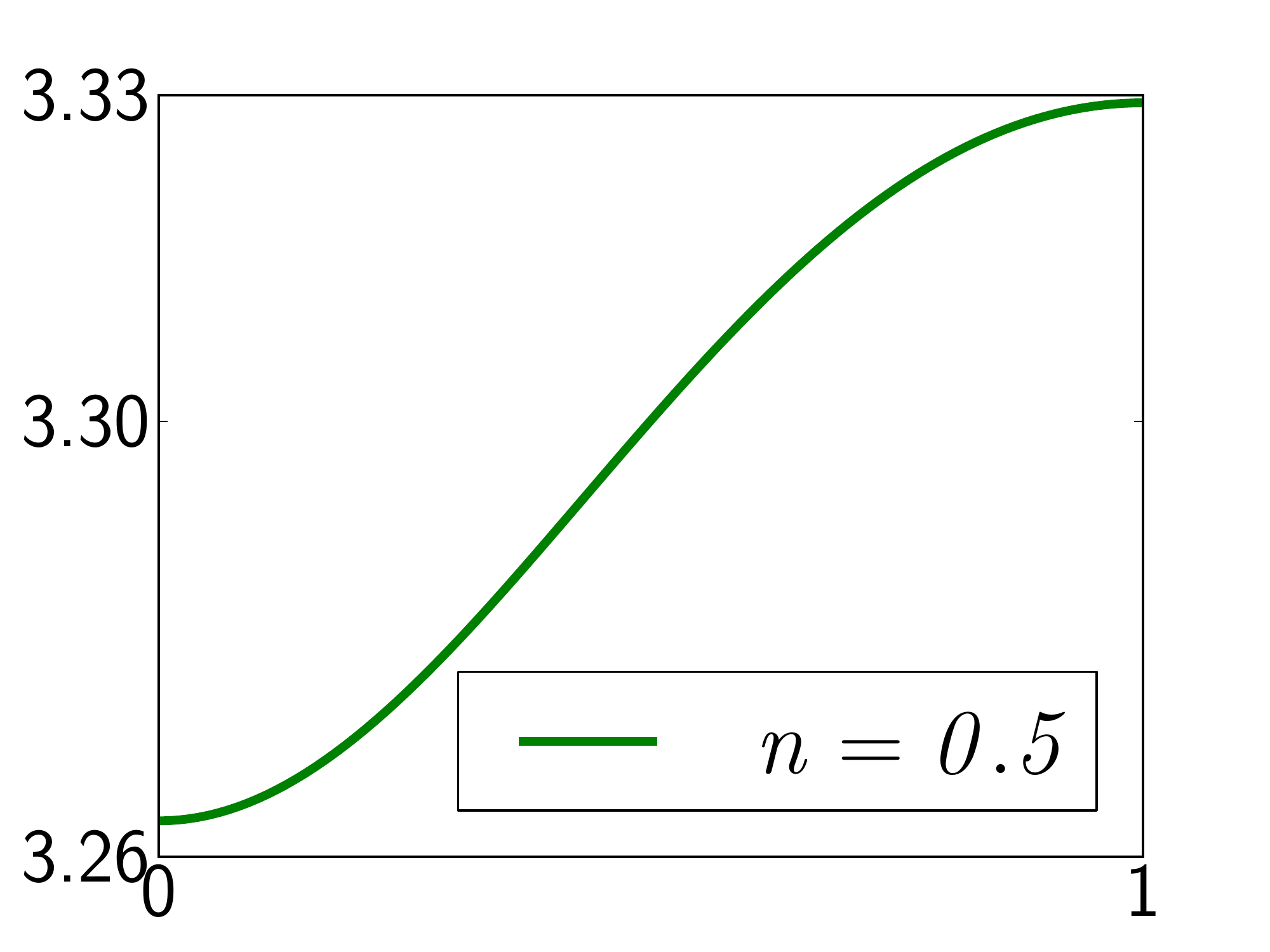}}}}

\caption{\label{fig:nothalf}Entanglement entropy $S(N,g)$ is plotted as a function of $g$ for $N=16$ sites, for various values of the number density of electrons $n$. The inset shows an enlarged plot for quarter-filling. }
\end{figure}    

\begin{figure*}

\begin{tabular}{cccc}
\subfloat[\label{logvlogy1}]{\includegraphics[scale = 0.45]{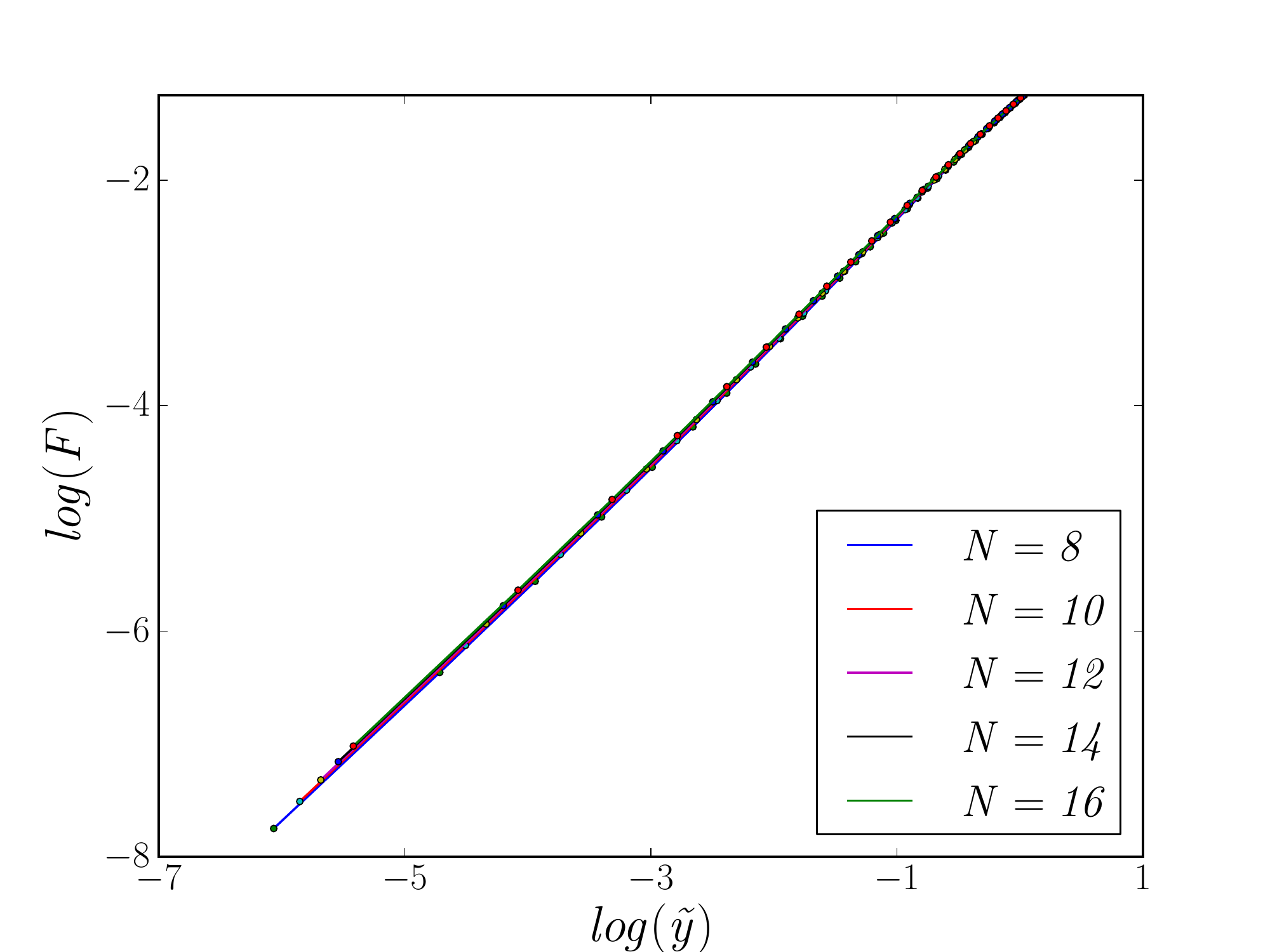}} &
\subfloat[\label{vE1}]{\includegraphics[scale=0.45]{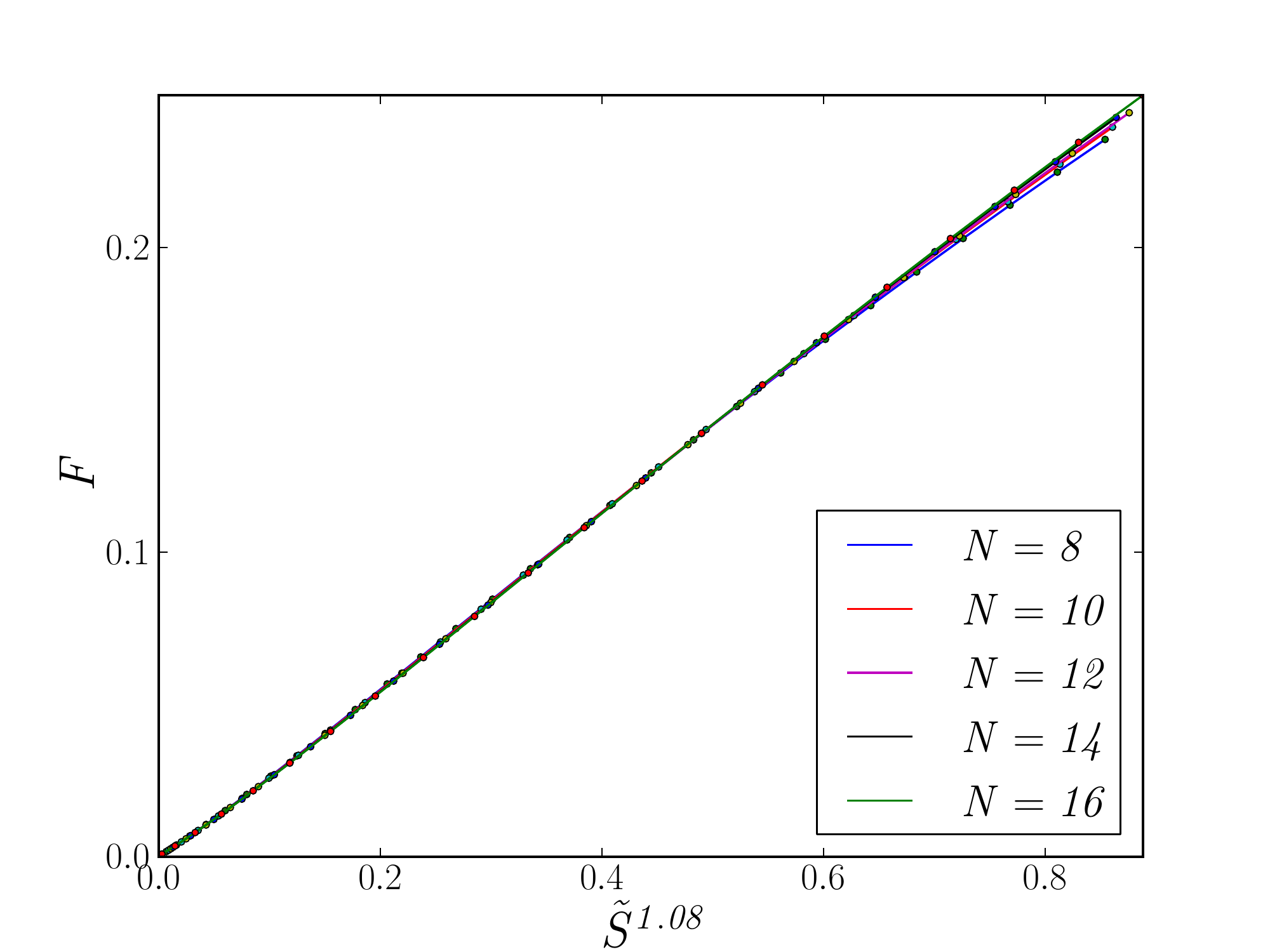}} & 

\end{tabular}
\caption{(a) Data collapse onto a straight line of slope $\simeq 1.08$ of log $F$ plotted against log $\tilde{y}$ for $\tilde{y} < 1$, $\tilde{y}=yZ(y)S(N,0)$. (b) $F$ is plotted with $\tilde{S}^{1.08}$ (Eq.\ref{E1}) for $\tilde{S}<1$ (and hence $\tilde{y}<1$) showing a data collapse onto a straight line passing through origin. This means the ansatz Eq. \ref{E(y)} is valid in this limit. The data collapse becomes slightly broad as $\tilde{S}$ approaches $1$. For $\tilde{S}>1$, there is no data collapse (not shown in figure), signifying Eq. \ref{E(y)} is not valid in this range.   } 
\end{figure*}

\section*{Away from half-filling}
Away from half-filling, the entanglement entropy for the non-interacting case, $g=1$, still shows a logarithmic divergence with central charge $c=2$, implying a metallic behaviour. However a decrease in the number of electrons causes a decrease in the entanglement entropy in the metallic state. Fig.\ref{fig:nothalf} shows the variation of entanglement entropy for $N=16$ sites for the case of quarter-filling and one-eighth filling. Analogous to the global entanglement \cite{A}, the entanglement entropy is seen to be  smaller in the strongly interacting limit than in the non-interacting limit. The effect of correlation projection factor $g$ decreases as number density decreases. This is because the average double occupancy itself changes by a smaller amount in these cases. For the half-filled case, the average double occupancy varies from $1/4$ for the metallic state to zero in the insulating state,  where as it changes from $1/{16}$ to zero for the  quarter-filled case, and from ${1}/{64}$  to zero for the one-eighth filling.  The entanglement entropy at $g=0$ for $n=1/2$ (quarter filling) is more than at $g=0$ for $n=1$ (half filling), as there are unoccupied sites (holes) in the quarter filling case which increase the entropy.

\section{\label{BFGS}Bipartite fluctuations in the half filled Gutzwiller state}
Bipartite fluctuations, i.e, fluctuations in a subsystem, have been seen to successfully describe quantum critical points \cite{Fluc}. In this section we study the bipartite fluctuations for half bipartition and their relation with the entanglement entropy in the half filled Gutzwiller state.
\section*{Number fluctuations in a subsystem} 

The question we now ask is how the fluctuation in the number of particles in one subsystem varies with the correlation factor $g$, and how it is related with the entanglement entropy. The fluctuation $F$ is characterised by the variance of the number of particles $N_A$ in subsystem $A$. This is given by :
\begin{equation}
F= <{N_A}^2> - {<N_A>}^2
\end{equation}
where $<\bullet> = \frac{Tr \hspace{4pt}\bullet\rho_A}{Tr \hspace{4pt}\rho_A} $, $\rho_A$ being the reduced density matrix of subsystem $A$. Using the symmetries of the system, the average number of particles in a subsystem can be easily seen to be $<N_A> = \frac{N}{2}$, for all values of $g$. The variance $F$ however depends on $g$.
It is very intuitive that the number fluctuation in the subsystem is directly related to conductivity of the system. For $g=0$, no fluctuations are allowed and hence the $F$ is zero. For $g=1$, maximum fluctuations are allowed and hence $F$ is maximum. Also, greater fluctuations are allowed for higher $N$. 

Since $F$ depends on $g$ and $N$, we now want to find if there is a scaling variable which is a function of $N$ and $g$ in terms of which $F$ will show a data collapse. We have done a similar thing for entanglement entropy previously, and we have the scaling variable $y$ already at hand. Since entanglement entropy is a measure of fluctuations in a system, we expect $c(N,g)$ and $F$ to be related. This means the scaling variable for $F$ should have some dependence on $y$. Since $F$ represents number fluctuations in a subsystem, we expect it to have a bearing on conductivity, and therefore expect $F$ to depend on the discontinuity  at the Fermi level. Motivated by these insights we find by inspection that a near data collapse is seen when $F$ is plotted as function of $\tilde{y}=yZ(y)S(N,0)$. The data collapse is especially good for $\tilde{y} < 1$. This fact is further strengthened by the log-log plot. When $log \hspace{4pt} v$ is plotted against $log \hspace{4pt} \tilde{y}$, for $\tilde{y} < 1$, there is very good data collapse onto a single straight line of slope $\simeq 1.08$ (Fig.\ref{logvlogy1}).

\begin{figure*}

\begin{tabular}{cccc}
\subfloat[\label{Ex}]{\includegraphics[scale = 0.45]{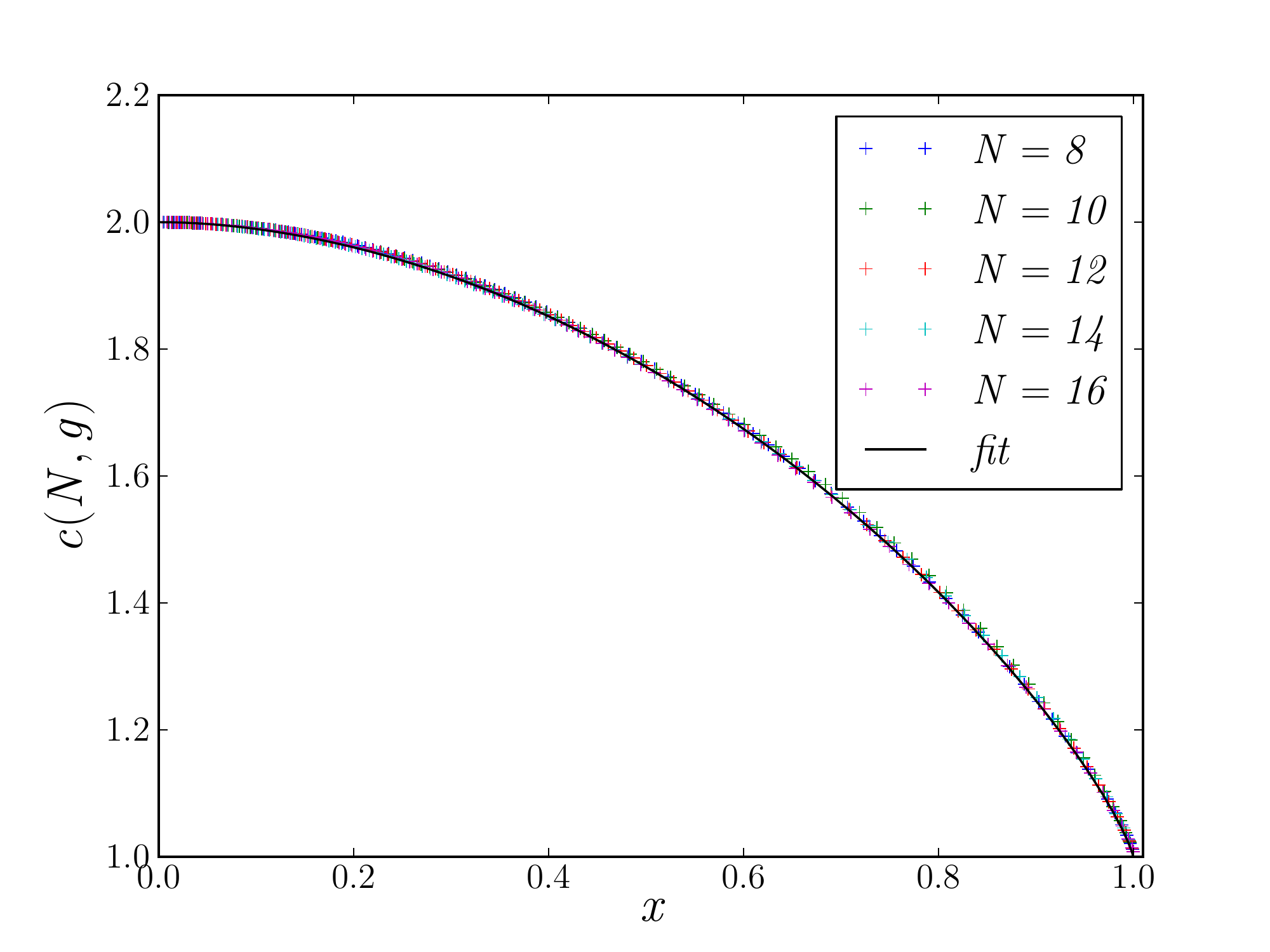}} & 
\subfloat[\label{dEdxx}]{\includegraphics[scale = 0.45]{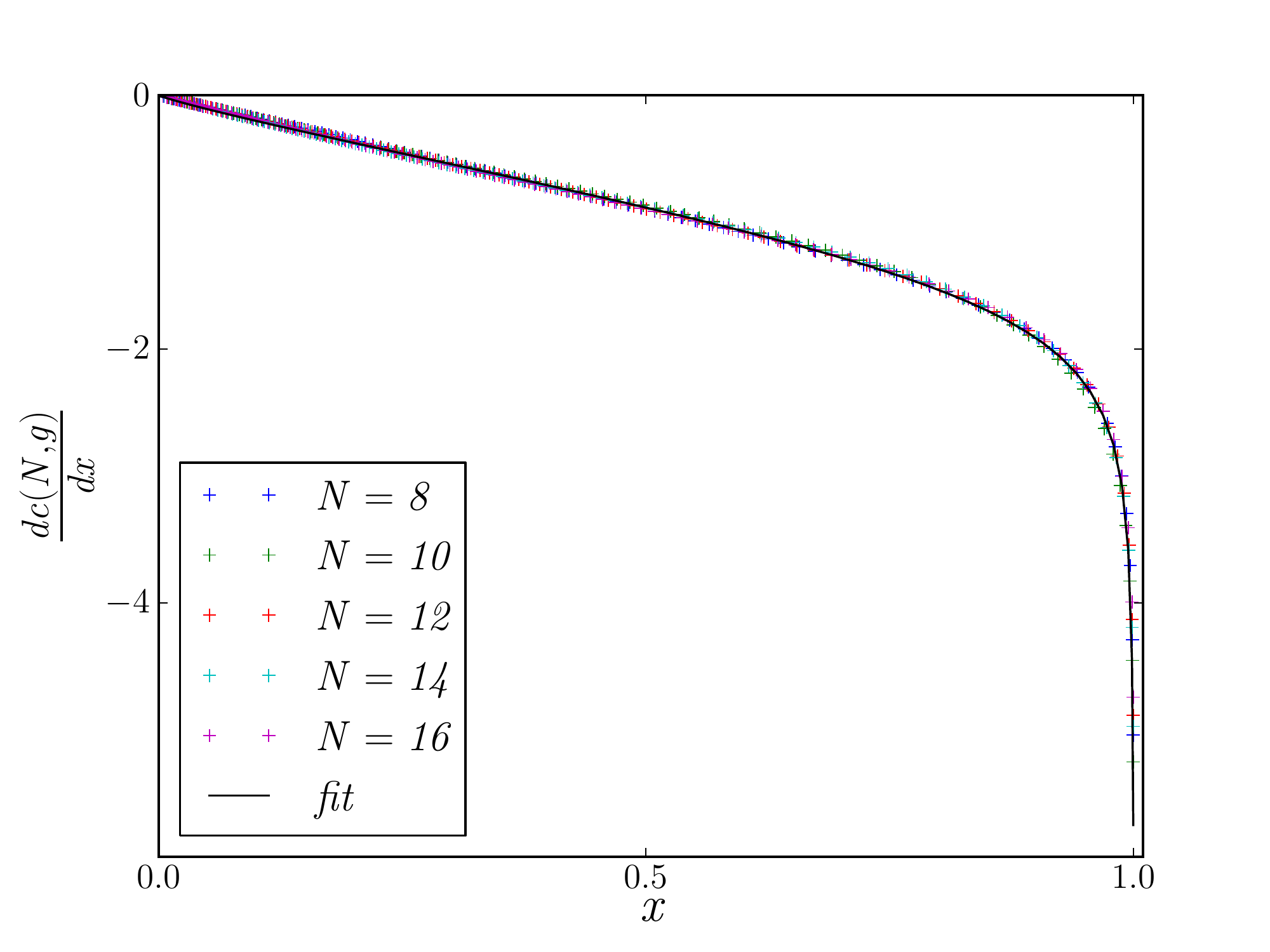}} &

\end{tabular}
\caption{\label{E}(a) Plot of $c(N,g)$ vs $x=\frac{F_{\uparrow\downarrow}}{F_{\uparrow\downarrow}(0)}\sqrt{\frac{F_{\uparrow}(0)}{F_{\uparrow}}}$ (as defined in Eq \ref{x}), showing a data collapse. The fit shows a plot of Eq \ref{f(x)} with $q=1.85$, $r=0.83$. (b) Plot of $\frac{dc(N,g)}{dx}$ vs $x$, as well as plot of derivative of Eq \ref{f(x)} with $q=1.85$, $r=0.83$. } 
\end{figure*}

At least in the region where the data collapse is good, i.e, when $\tilde{y} < 1$, we can now check whether our ansatz for $c(y)$ given in Eq \ref{E(y)} is valid. To check for this we calculate $\tilde{S}$, given by :
\begin{equation}
\label{E1}
\tilde{S} = (log(\beta)-log(\alpha-S(N,g)))S(N,0)/\kappa
\end{equation}
If our ansatz Eq. \ref{E(y)} is valid, $\tilde{S}=\tilde{y}$. Therefore, a plot of $F$ against $\tilde{S}^{1.08}$ should be a data collapsed straight line passing through origin for $\tilde{S}<1$. This is exactly what is seen. (Fig \ref{vE1}). This means the ansatz Eq. \ref{E(y)} is valid in this limit. However as $\tilde{S}$ (and hence $\tilde{y}$) approaches $1$, the data collapse becomes slightly worse. For $\tilde{S}>1$ (not shown in Fig \ref{vE1}), there is no data collapse at all, proving that the ansatz is not valid in this range. 
Thus the range of $y$ for which Eq. \ref{E(y)} is valid is :
\begin{equation}
y<\frac{1+2\sqrt{S(N,0)}}{4S(N,0) - 1} = y_{l}
\end{equation}
We have seen from in the previous section that when Eq. \ref{E(y)} is valid, metal-insulator crossover point is at $y=y_0 \sim 0.24$. Thus Eq. \ref{E(y)} will continue to describe the metal-insulator crossover point as long as $y_l>y_0$. This condition yields a corresponding range of $N$, which is :
\begin{equation}
N<2^{18} \sim 10^5
\end{equation}  

So, for system sizes in this range we have located the metal-insulator crossover point. This range of system sizes is physically realized in nanochains. Nanochains are typically $\sim 1 \mu m$ long with lattice spacing $\sim 0.1 nm $, thereby having $N \sim 10^4$. Thus our results provide an understanding of metal-insulator transition in nanochains of strongly correlated materials.

The appearance of $S(N,0)$ in the scaling relation for $F$ is also to be noted here. $S(N,0)$ is the entanglement entropy of the system when no number fluctuations are allowed in the subsystem. Thus it is a measure of all fluctuations other than number fluctuations. This seems to provide a scale in terms of which number fluctuations are to be measured. This also provides an intuitive reason for $c(N,g)$ to capture the metal-insulator behaviour. By rescaling by $S(N,0)$ to calculate $c(N,g)$ we isolate the contribution of number fluctuations to the entanglement, which is the primary reason for conductivity of the system. When number fluctuations are not allowed, there can be spin flips preserving the number of particles in the subsystem. $S(N,0)$, therefore captures these fluctuations, which are always present for any value of $g$. Changing $g$ does not affect these fluctuations and therefore the contribution to entanglement entropy from this is the same throughout. So $S(N,0)$ appears as a multiplicative factor in the expression for entanglement entropy. With change in $g$, only the central charge $c(N,g)$ changes.

\section*{Spin fluctuations in a subsystem}

Even when there is no number fluctuation in the subsystem, there can be spin fluctuations in the subsystem. Spin fluctuations come in two kinds : fluctuations related to spins of same kind ($F_{\uparrow}$), and fluctuations related to spins of different kind ($F_{\uparrow\downarrow}$). But these two are related to number fluctuations by the relation : $F=2(F_{\uparrow} + F_{\uparrow\downarrow})$, where, $F_{\uparrow}$ and $F_{\uparrow\downarrow}$ are given as:
\begin{align}
F_{\uparrow} & = <{N_{A\uparrow}}^2> - <N_{A\uparrow} >^2 \\
F_{\uparrow\downarrow} & = <N_{A\uparrow} N_{A\downarrow} > - <N_{A\uparrow} ><N_{A\downarrow} >
\end{align} 
where $N_{A\uparrow}$ and $N_{A\downarrow}$ are the number of up and down spins in subsystem respectively.

We want to know what bearing $F_{\uparrow}$ and $F_{\uparrow\downarrow}$ have on $c(N,g)$. Since in $c(N,g)$, the contribution from the $g=0$ correlations is factored out, we look for the dependence of $c(N,g)$ on $\frac{F_{\uparrow\downarrow}}{F_{\uparrow\downarrow}(0)}$ and $\frac{F_{\uparrow}}{F_{\uparrow}(0)}$, where $F_{\uparrow}(0)$ and $F_{\uparrow\downarrow}(0)$ are values of $F_{\uparrow}$ and $F_{\uparrow\downarrow}$ at $g=0$. By inspection it is seen that a very good data collapse is obtained when $c(N,g)$ is plotted as a function of : 
\begin{equation}
\label{x}
x=\frac{F_{\uparrow\downarrow}}{F_{\uparrow\downarrow}(0)}\sqrt{\frac{F_{\uparrow}(0)}{F_{\uparrow}}}
\end{equation}

The plots showing data collapse are given in Fig \ref{E}.   At $g=0$, $x=1$, $c(N,g)=1$, and at $g=1$, $x=0$, $c(N,g)=2$. The plot of $c(N,g)$ vs $x$ is bounded by these values for any $N$. Fig \ref{dEdxx} shows plot of derivative of $c(N,g)$ vs $x$. We see that the derivative is zero at $x=0$ but diverges to $-\infty$ at $x=1$. Let us assume that $c(N,g)$ can be described by a polynomial function $f(x)$. Then the fact that the derivative diverges while the function itself is well defined at $x=1$ suggests there must be a term of the form $(1-x)^r$ with $0<r<1$. The derivative is zero at $x=0$. This means there must be a linear part in $f(x)$ which, when differentiated, cancels the contribution of $(1-x)^r$ term in the derivative of $f(x)$ at $x=0$. Let us assume the remaining terms in $f(x)$ does not involve a constant term. Then the fact that at $x=0$, $c(N,g)=2$ implies the coefficient of $(1-x)^r$ term is $2$. Then the coefficient of the linear part must be $2r$ so as to make the derivative zero at $x=0$. Finally, let us assume that the remaining term in $f(x)$ is of the form $x^q$. Since $c(N,g)$ vs $x$ plot is decreasing and concave for all $x$, $q>1$. The fact that at $x=1$, $c(N,g)=1$, then implies that the coefficient of $x^q$ term is $(1-2r)$. Hence we have an equation of the form :
\begin{equation}
\label{f(x)}
c(N,g) = f(x) = (1-2r)x^q + 2rx + 2(1-x)^r
\end{equation}  

\begin{figure}[t]
\includegraphics[scale = 0.45]{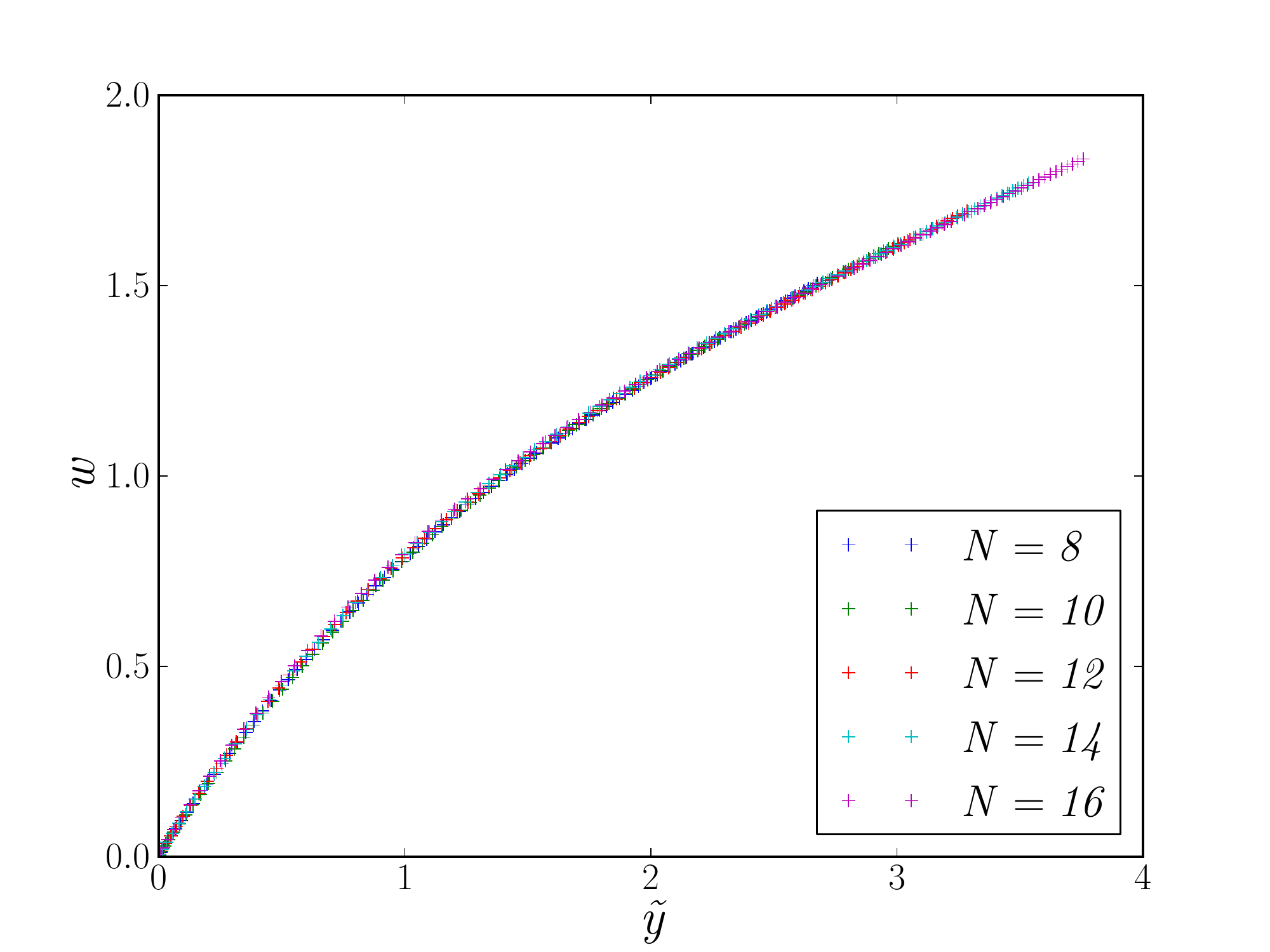}
\caption{\label{wy1} Plot of $w$, defined in Eq \ref{w}, with $\tilde{y}$. Remarkably, the plot seems to show a data collapse for all values of $\tilde{y}$, but closer inspection reveals that the data collapse is good for small $\tilde{y}$, as expected. Nevertheless, such data collapse suggests that our choice of $y=N^{1/3}g$ is to be questioned. However it does quite well when $\tilde{y}<1$. } 
\end{figure}

When this equation is fitted with $q$ and $r$ as parameters, quite a good fit is obtained and the values come out to be $q\simeq 1.85$, $r\simeq 0.83$. The fits are also shown in Fig \ref{Ex} and \ref{dEdxx}. The fractional values of $q$ and $r$ means that $f(x)$ is only real in the range $0\leq x \leq 1$, which is exactly the accessible range in the present case.

Next we seek to consolidate Eq \ref{f(x)} with our previous findings. We have found that Eq \ref{E(y)} is valid for $\tilde{y}<1$. This limit implies small values of $g$. Thus Eq \ref{f(x)} must reproduce Eq \ref{E(y)} as $g\rightarrow0$, i.e, as $x\rightarrow 1$. The term in $f(x)$ that varies most slowly with $x$ is $2(1-x)^r$. So, as $x\rightarrow 1$, we have the following relation :
\begin{equation}
\label{apprx}
c(N,g) = f(x) \simeq 1+2(1-x)^r \simeq 1 + 1.75 y Z(y)
\end{equation}
where the last relation has been obtained by expanding Eq \ref{E(y)} for small $yZ(y)$. Thus, in this limit, :
\begin{equation}
\label{w}
\tilde{y} = yZ(y)S(N,0) \sim (1-x)^r S(N,0) = w \hspace{8pt} (say) 
\end{equation}
We have multiplied both sides by $S(N,0)$ because $\tilde{y}$ is the variable of interest here.

Fig \ref{wy1} shows plots of $w$ vs $\tilde{y}$. Remarkably, there seems to be a data collapse for all values of $\tilde{y}$, even though, based on Eq \ref{w}, we would have expected that to occur only for small $\tilde{y}$. However, closer inspection shows that the data collapse is not good for the entire range of $\tilde{y}$. It is good only for small $\tilde{y}$, as expected. Nevertheless, the fact that there is a near data collapse even for the part where the dependence is not approximately linear suggests that $x$ is not just a function of $y$. There is an additional $N$ dependence coming through $S(N,0)$. Since $c(N,g)=f(x)$, this in turn suggests that our initial choice of $y=N^{1/3}g$ is not good for the entire range of values. It is only good for small $y$, and is seen to hold for $\tilde{y}<1$ where Eq \ref{E(y)} holds to a good approximation. Our prediction about the metal-insulator crossover point holds in this region. Only Eq \ref{f(x)}, the equation relating central charge to spin fluctuations in subsystem, is valid for the entire range of $g$.

\section{\label{MIT}Metal-insulator crossover in nanochains}
Since we are working with one dimensional finite-size systems we are essentially dealing with nanochains. From the crossover behaviour of the entanglement entropy, we may infer that metal-insulator transitions can occur in nanochains. At half-filling, the crossover is marked by the scaling variable (the location of the peak in Fig.3) which combines the system size and the correlation factor, given by
\begin{equation}
y=N^{\frac{1}{3}}g \simeq 0.24 \label{crossover}.
\end{equation}
Therefore, for a fixed finite size,  the metal-insulator transition can occur at a finite value of $g\simeq 0.24/N^{1/3}$. However, in the thermodynamic limit, it can occur only at $g=0$, as is well known. Conversely, a metal-insulator transition can also occur by changing the size of the system, for a fixed correlation factor. Thus,  both an interaction-induced, as well as a size-induced metal-insulator transition can be explained from our study of entanglement in the Gutzwiller state.

Let us now see how Eq.\ref{crossover} compares with the experiments and the known results. However,  caution should be exercised  in applying the above to real systems, as we are talking of a one-dimensional variational state in comparison with real three-dimensional materials.  Also, Gutzwiller state is not applicable to every material. One material it has had considerable success explaining is Ni. The Gutzwiller state has been quite successful in explaining band structure from ARPES results for Ni \cite{Ni}. So we expect our results might be observable in Ni nano-chains. Ni has two possible ground state electron configurations with nearly same energy, i.e. $\rm{[Ar]4s^23d^8}$ and $\rm{[Ar]4s^13d^9}$. In the spirit of Hubbard's original paper \cite{H}, we look at the d-electrons only, even though we have assumed s-band electrons for mathematical ease. The second configuration of Ni has one hole in the d-orbital. Since at half-filling the Hubbard model and the Gutzwiller state are known to have particle hole symmetry, this case can be roughly thought of as a half-filled system. Therefore, we expect the results at half-filling, that we discussed above, to hold for Ni to some extent.

The on-site correlation factor $g$ is related to Hubbard parameters $t$ and $U$ through Eq.\ref{gU}. For Ni, the bandwidth measurements have yielded the experimental
value of $t$, and the interaction strength $U$ has been obtained by fitting the theoretical results to band structure measurements. Values of $t$ have been reported for various bands of Ni\cite{Ni}. Since, we are considering only d-electrons but have calculated assuming s-band electrons, we will consider the value of $t$ for $dd\sigma$ band which retains the symmetry of the atomic s-orbital. Also, $dd\sigma$ band has been reported to have the greatest contribution to the to the 3d band width.  For this case, the reported value of $t\sim0.5eV$. The value of $U\sim10eV$ has been reported to reproduce the experimental 3d band width. Therefore, within this approximation, we estimate $t/U \sim 0.05$ for Ni. We assume in this calculation that $g$, and hence $t/U$, is independent of system size. However, as dimension decreases, the value of $t$ which depends on the overlap of nearby orbitals is expected to remain almost the same, but the value of $U$ is expected to increase due to decrease in screening. Thus in one dimension, the value of $t/U$ is expected to be lower than the above estimate.

In a recent experiment\cite{nanochain}, Ni nanochains of length $2-3\mu m$ have been fabricated and have been reported to be antiferromagnetic and insulating. Typical lattice parameter for Ni is $\sim 0.3nm$. Therefore,  we can estimate the number of sites for these Ni nanochains to be  $N\sim 10^4 < 10^5$, thus our description of metal-insulator crossover holds. With this value of $N$, combining Eq.\ref{crossover} and Eq.\ref{gU} gives a metal-insulator crossover value at $t/U \sim 0.05$ which is exactly the value we got by rough estimation from known results in three dimensions. Since in one dimension, the value is expected to be smaller than this value, the nanochain is expected to be insulating, as is experimentally seen.

However, the estimation of $t/U$ from known three dimensional results was quite rough. Hence, based on the above calculations, all that can be concluded is that Eq.\ref{gU} gives at least `in the ballpark' results. Even though this is only a rough approximation, such correspondence suggests that longer Ni nano-chains will be pushed towards being metallic. If this trait is seen in experiments, it can be taken as verification of our results. This is more so because the exact solution of Hubbard model in one dimension in thermodynamic limit predicts the opposite, viz. an insulating state.

\section{\label{conclusion} Conclusion}

In conclusion, we have investigated the entanglement entropy and bipartite fluctuations  for the Gutzwiller state in one dimension, and the crossover behaviour at the metal-insulator transition in finite systems.

We have shown that the entanglement entropy $S(N,g)$ for the half filled Gutzwiller state shows logarithmic divergence for all values of the correlation factor $g$. This is because the correlation between same kind of spins always have an infinite correlation length, as we have shown by calculating correlation functions by variational Monte Carlo. Only the coefficient of the logarithmic term varies with $g$, capturing the effect of correlation between opposite spins which decreases with $g$. In $g=1$ and $g=0$ limits, the coefficient becomes the central charge of the underlying CFT even for small system sizes $N$. For intermediate values, the coefficient is interpreted is central charge along with some finite size corrections, $c(N,g)$. Plots of derivative of $c(N,g)$ with $g$ show a peak which correspond to the value of $g$ where finite size corrections are maximum. Based on correlation length arguments, this value of $g$ is interpreted as the insulator to metal crossover point. An approximate scaling relation for $c(N,g)$ is found in terms of $y=N^\frac{1}{3}g$ that does not hold for all values of $N$ and $g$, but describes the peak at least for small values of $N$. This scaling relation depends on $Z(y)$, where $Z(g)$ gives the discontinuity of the momentum distribution function at the Fermi level. In terms of $y$, the insulator to metal crossover point is at $y=y_0 \simeq 0.24$. We have also found approximate scaling law for bipartite number fluctuations in half filled Gutzwiller state, and from this result we have shown that $y_0$ gives the crossover point for $N<2^{18} \sim 10^5$. We have also found an expression for  $c(N,g)$ in terms of bipartite spin fluctuations.
          
Away from half-filling, in the non-interacting ($g=1$) case entanglement still shows logarithmic divergence with central charge $c=2$, showing metallic behaviour. However, the entanglement entropy decreases with a decrease in number density of electrons in the metallic limit. Also, the effect of $g$ reduces. This is attributed to the fact that the difference between the average double occupancy of non-interacting and strongly interacting limits decreases with a decrease in the number density. The entanglement entropy varies non-monotonically with the number density of electrons in the strongly-interacting limit.

From our semi-quantitative analysis of entanglement in Gutzwiller state in one dimension, we find that the Gutzwiller state at half-filling can show a metal-insulator transition at a finite value of the on-site correlation factor $g$ and for a finite size $N$. Our expression for the crossover point shows that both the interaction-induced and the size-induced metal-insulator transitions can be described by the Gutzwiller state. Finally, we have shown that our results show some correspondence with a recent experiment on Ni nanochains. Such correspondence predicts that longer Ni nanochains will be pushed towards being metallic, which is in contrast with the exact solution of one dimensional Hubbard model, which predicts an insulating state in the thermodynamic limit.

\end{document}